\documentclass[prl,twocolumn,superscriptaddress,fixfloat]{revtex4}
\usepackage{amsmath}
\usepackage{amssymb}
\usepackage{graphicx}
\begin{document}

\title{The emergence of coherent magnetic excitations in the pseudogap phase of
La$_{2-x}$Sr$_{x}$CuO$_{4}$}

\author{O.J. Lipscombe}
\affiliation{H.H. Wills Physics Laboratory, University of Bristol, Tyndall Ave., Bristol,
BS8 1TL, UK}
\author{B. Vignolle}
\affiliation{H.H. Wills Physics Laboratory, University of Bristol, Tyndall Ave., Bristol, BS8 1TL, UK}
\author{T.G. Perring}
\affiliation{ISIS Facility, Rutherford Appleton Laboratory, Chilton, Didcot, Oxfordshire
OX11 0QX, United Kingdom}
\author{C. D. Frost}
\affiliation{ISIS Facility, Rutherford Appleton Laboratory, Chilton, Didcot, Oxfordshire
OX11 0QX, United Kingdom}
\author{S.M. Hayden}
\affiliation{H.H. Wills Physics Laboratory, University of Bristol, Tyndall Ave., Bristol,
BS8 1TL, UK}

\begin{abstract}
We use inelastic neutron scattering to measure the magnetic excitations in the underdoped superconductor La$_{2-x}$Sr$_{x}$CuO$_{4}$ (x=0.085, $T_c$=22~K) over energy and temperatures ranges $5<E<200$~meV and  $5<T<300$~K respectively. At high temperature ($T=300$~K), we observe strongly damped excitations with a characteristic energy scale $\Gamma \approx 50$~meV. As the temperature is lowered to $T=30$~K, and we move into the pseudogap state, the magnetic excitations become highly structured in energy and momentum below about 60~meV. This change appears to be associated with the development of the pseudogap in the electronic excitations.
\end{abstract}

\pacs{74.72.Dn, 74.25.Ha, 75.40.Gb, 78.70.Nx}

\maketitle

In addition to their high transition temperatures, the cuprate superconductors have many intriguing properties.  For example, the electronic properties change dramatically with doping.  In the  underdoped region of the phase diagram, a `pseudogap' develops at temperatures well above the superconducting transition temperature $T_c$.  The pseudogap is reminiscent of a loss of low-energy electronic spectral weight and had been observed in many electronic properties \cite{Timusk1999,Tallon2001} including nuclear magnetic resonance (NMR), Angle Resolved Photoemission Spectroscopy (ARPES), Raman spectroscopy, optical conductivity and scanning tunneling microscopy (STM).  For overdoped compositions, the electronic gap appears to form at $T_c$ as in conventional superconductors.  Many believe that the pseudogap is at the heart of high-$T_c$ superconductivity phenomenon and that it is key to understanding the high-$T_c$ mechanism.

There are two main scenarios which purport to describe the pseudogap.  The first is that it is due to a competing ordered phase such as orbital currents or charge-density-wave ordering \cite{Kyung2006a,Varma2007a,Schmalian2008a}.  In this `two-gap' scenario, the superconducting gap vanishes above $T_c$ leaving the gap due to the competing order.  In the alternative `one-gap' scenario, the pseudogap is a vestige of the superconducting gap, with fluctuations destroying long-range coherence for $T>T_c$ \cite{Emery1995a,Norman1998a,Chubukov2007a}. In other words, the pseudogap is a precursor to the superconductivity.  Charge spectroscopies show evidence for low-temperature gaps or suppression in spectral weight in the energy range 40--100~meV, which develop near the `pseudogap temperature', $T^{\star}$ \cite{Timusk1999}.  In contrast, large gaps or a suppression of spectral weight over a corresponding large energy range have \emph{not} been observed in the collective spin excitations for underdoped compositions \cite{Bourges1997,Dai1999}.  In this letter, we use inelastic neutron scattering (INS) to determine the effect of the pseudogap on the spin excitations in underdoped La$_{2-x}$Sr$_{x}$CuO$_{4}$ (LSCO). Measurements were made over wider ranges in energy ($5<E<200$~meV) and temperature ($5<T<300$~K) than previous studies \cite{Keimer1991,Hayden1991a,Lee2000a,Kofu2007a}.  We find that as the pseudogap state is developed by lowering the temperature, there is a dramatic emergence of coherent magnetic excitations below the pseudogap energy $2\Delta^{\star}$.

Our experiments were performed on underdoped La$_{2-x}$Sr$_{x}$CuO$_{4}$ with $x$=$0.085 \pm 0.005$ and $T_c$=22~K. This composition has a pseudogap temperature $T^{\star}$ $\approx$\ 400--500~K. Four single crystals with a total mass of 40.5~g were co-aligned with a total mosaic of 1.5$^\circ$. The crystals were grown by a traveling-solvent floating-zone technique \cite{Komiya2002a} and annealed with one bar of oxygen for one week at 800$^\circ$C. The Sr stoichiometry was measured with SEM-EDX and ICP-AES to be $x=0.085 \pm 0.005$. Magnetization measurements indicate that $T_c \mathrm{(onset)}= 22$~K.

INS probes the energy and wavevector dependence of $\chi^{\prime
\prime}(\textbf{q},\omega)$.  The magnetic cross section is given by
\begin{equation}
\label{Eq:cross_sect} \frac{d^2\sigma}{d\Omega \, dE} = \frac{2(\gamma
r_{\text{e}})^2}{\pi g^{2} \mu^{2}_{\rm B}} \frac{k_f}{k_i} \left| F({\bf Q})\right|^2
\frac{\chi^{\prime\prime}({\bf q},\hbar\omega)}{1-\exp(-\hbar\omega/kT)},
\end{equation}
where $(\gamma r_{\text{e}})^2$=0.2905 barn sr$^{-1}$, ${\bf k}_{i}$ and ${\bf k}_{f}$ are the incident and final neutron wavevectors and $|F({\bf Q})|^2$ is the anisotropic magnetic form factor for a Cu$^{2+}$ $d_{x^{\scriptstyle 2}-y^{\scriptstyle 2}}$ orbital. Data were placed on an absolute scale using a vanadium standard.
We use the reciprocal lattice of the high-temperature tetragonal structure of La$_{2-x}$Sr$_{x}$CuO$_{4}$ ($x=0.085 $) to label wavevectors $\mathbf{Q}=h\mathbf{a}^{\star}+k\mathbf{b}^{\star}+l\mathbf{c}^{\star}$ and we will usually quote only the in-plane components $(h,k)$. In this notation, La$_{2}$CuO$_{4}$ exhibits antiferromagnetic order with an ordering vector of $(1/2,1/2)$. Our INS experiments were performed on the MAPS instrument at the ISIS spallation source. MAPS is a direct-geometry time-of-flight chopper spectrometer with position-sensitive detectors. This allows a large region of reciprocal space to be sampled using a single setting with a given incident energy $E_i$. In order to identify and minimize phonon contamination of our results, we collected data for six different $E_i$'s and utilized the procedure developed in Ref. \cite{Lipscombe2007a}.

\begin{figure*}
\begin{center}
\includegraphics[width=0.98\linewidth]{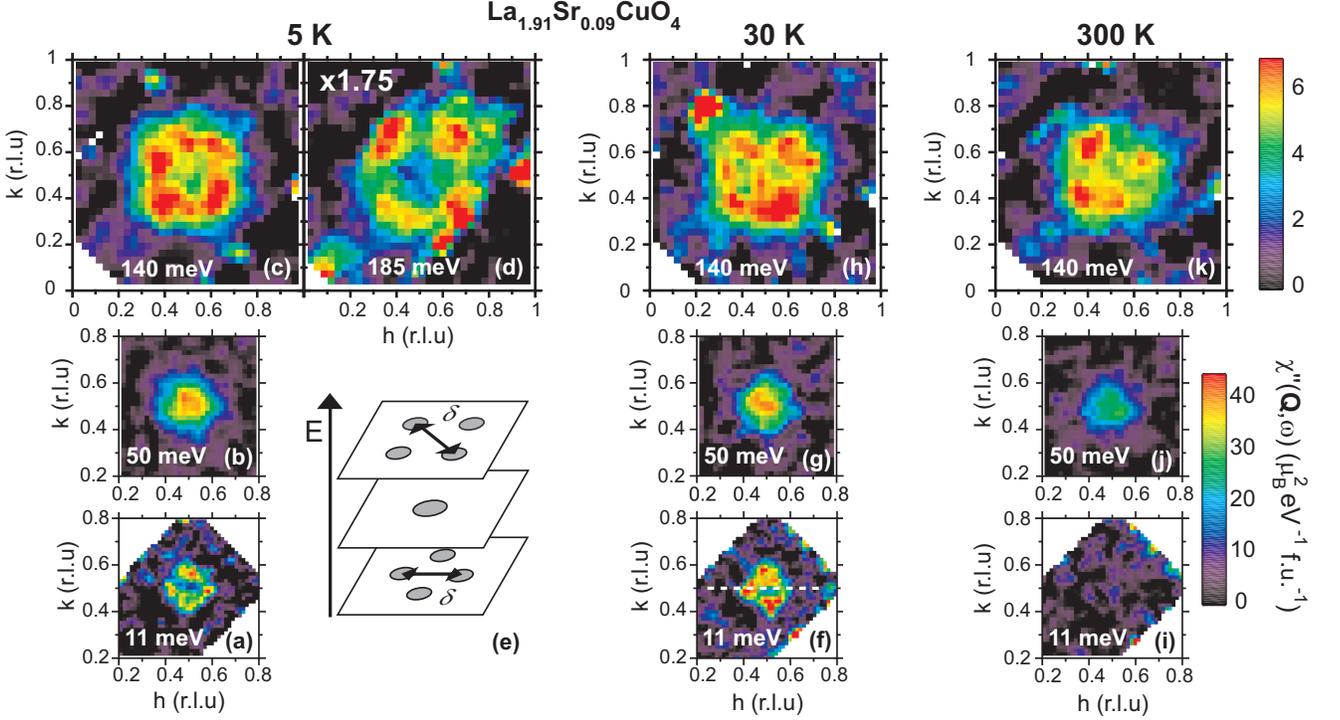}
\end{center}\caption{(Color online) Constant-$E$ slices of data converted to $\chi^{\prime\prime}(\mathbf{q},\omega)$ using Eq.~\ref{Eq:cross_sect}. (a)-(d) show the response at 5 K for increasing energy, (f)-(h) at 30 K and (i)-(k) at 300 K. Ranges of integration were $11 \pm 1$, $50 \pm 5$, $140 \pm 10$ and $185 \pm 15$~meV, using $E_i$=55, 240, 240, 240 meV respectively.  (e) shows the schematic evolution of the pattern with energy.}
\label{Fig:slices}
\end{figure*}

\begin{figure*}
\begin{center}
\includegraphics[width=0.98\linewidth]{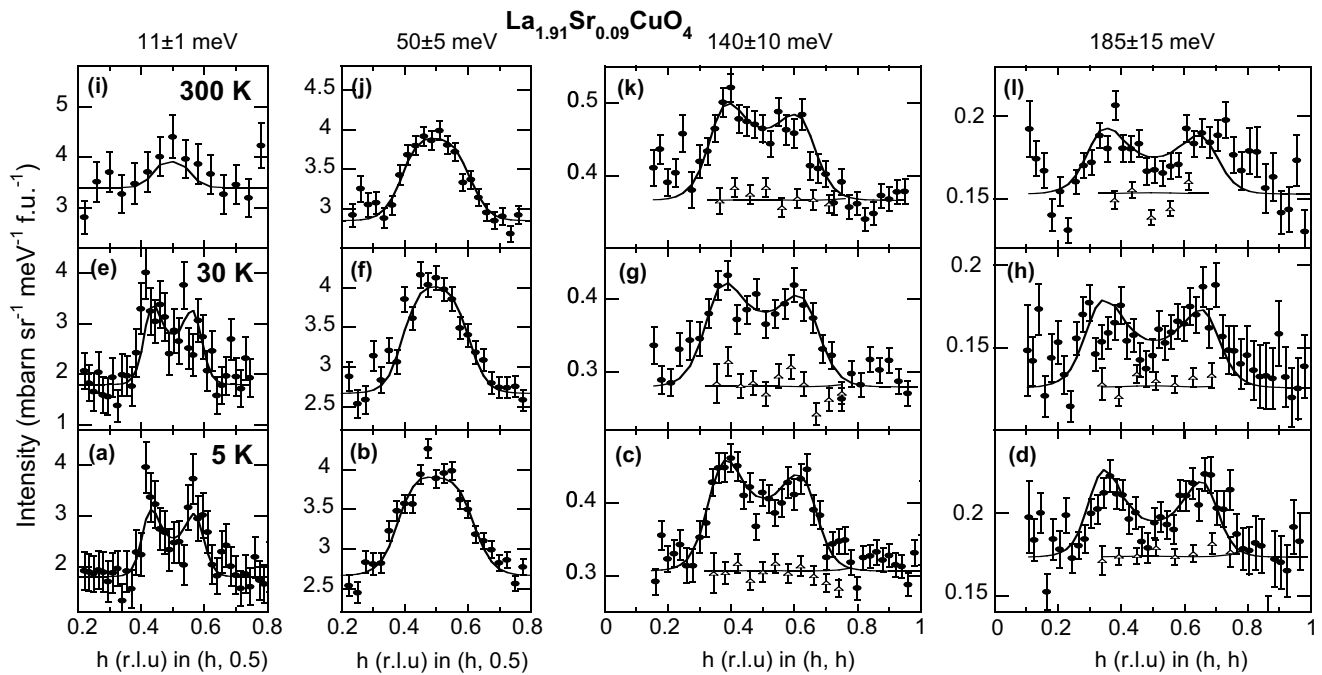}
\end{center}\caption{Constant-$E$ cuts through raw data used to produce the plots in Fig.~\ref{Fig:slices}. Data have not been corrected for the Bose or form factors in Eq.~\ref{Eq:cross_sect}.  Solid lines are fits of Eq.~\ref{Eq:Sato_Maki} convolved with the instrument resolution.} \label{Fig:cuts}
\end{figure*}

Fig.~\ref{Fig:slices} shows typical $\mathbf{q}$-dependent images of $\chi^{\prime\prime}(\textbf{q},\omega)$ for various energies and temperatures.  We first consider the response in the superconducting state for our sample of La$_{2-x}$Sr$_{x}$CuO$_{4}$ ($x$=0.085, $T_c$=22~K).  At low energy, $E$=11~meV, and temperature, $T$=5~K [Fig.~\ref{Fig:slices}(a)], we observe the well known four-peaked structure in $\mathbf{q}$ \cite{Cheong1991}, with  $\chi^{\prime\prime}(\mathbf{q},\omega)$  peaked at the incommensurate positions $(1/2,1/2 \pm \delta)$ and $(1/2 \pm \delta,1/2)$.  As for optimally doped LSCO, the excitations disperse strongly with energy \cite{Vignolle2007a}.  For $E$=50~meV [Fig.~\ref{Fig:slices}(b)], they are peaked at (1/2,1/2) and at $E$=140~meV [Fig.~\ref{Fig:slices}(c)] and 185 meV [Fig.~\ref{Fig:slices}(d)] they are peaked at $(1/2 \pm \delta/\sqrt{2}, 1/2 \pm \delta/\sqrt{2})$ and $(1/2 \mp \delta/\sqrt{2}, 1/2 \mp \delta/\sqrt{2})$. On raising the temperature, the most striking change occurs between 30~K and 300~K.  The low energy response is strongly suppressed for $E$=11~meV [Fig.~\ref{Fig:slices}(i)] and the suppression also occurs at 50 meV [Fig.~\ref{Fig:slices}(j)]. Fig.~\ref{Fig:cuts} shows cuts through the data in Fig.~\ref{Fig:slices}.

In order to make a quantitative analysis of our data, we fitted 2-D slices, such as those in Fig.~\ref{Fig:slices}, to a modified Lorentzian function \cite{Vignolle2007a}:
\begin{equation}
\label{Eq:Sato_Maki} \chi^{\prime\prime}({\mathbf q},\omega)=\chi_\delta(\omega)
\frac{\kappa^4(\omega)} {[\kappa^2(\omega)+R(\mathbf{q})]^2}
\end{equation}
with
\begin{equation*}
R(\mathbf{q})=\frac{\left[(h-\frac{1}{2})^2+(k-\frac{1}{2})^2-\delta^2 \right]^2+\lambda (h-\frac{1}{2})^2 (k-\frac{1}{2})^2} {4 \delta^2},
\end{equation*}
where the position of the four peaks is determined by $\delta$, $\kappa$ is an inverse correlation length (peak width), and $\lambda$ controls the shape of the pattern ($\lambda$=4 yields four distinct peaks and $\lambda$=0 a pattern with circular symmetry \cite{Vignolle2007a}). At higher energies, a $\pi/4$ rotated version of this function was used with peaks at $(1/2 \pm \delta/\sqrt{2}, 1/2 \pm \delta/\sqrt{2})$ and $(1/2 \mp \delta/\sqrt{2}, 1/2 \mp \delta/\sqrt{2})$. This phenomenological response function provides a good description of the data at all energies. We have expressed the strength of the magnetic response in terms of the wavevector-averaged or local susceptibility $\chi^{\prime\prime}(\omega)=\int \chi^{\prime\prime}(\mathbf{q},\omega) \; d^{3}q/\int
d^{3}q$ determined from the fitted $\chi^{\prime\prime}(\mathbf{q},\omega)$. The local susceptibility indicates the overall strength of the magnetic excitations for a given energy.

\begin{figure*}
\begin{center}
\includegraphics[width=0.95\linewidth]{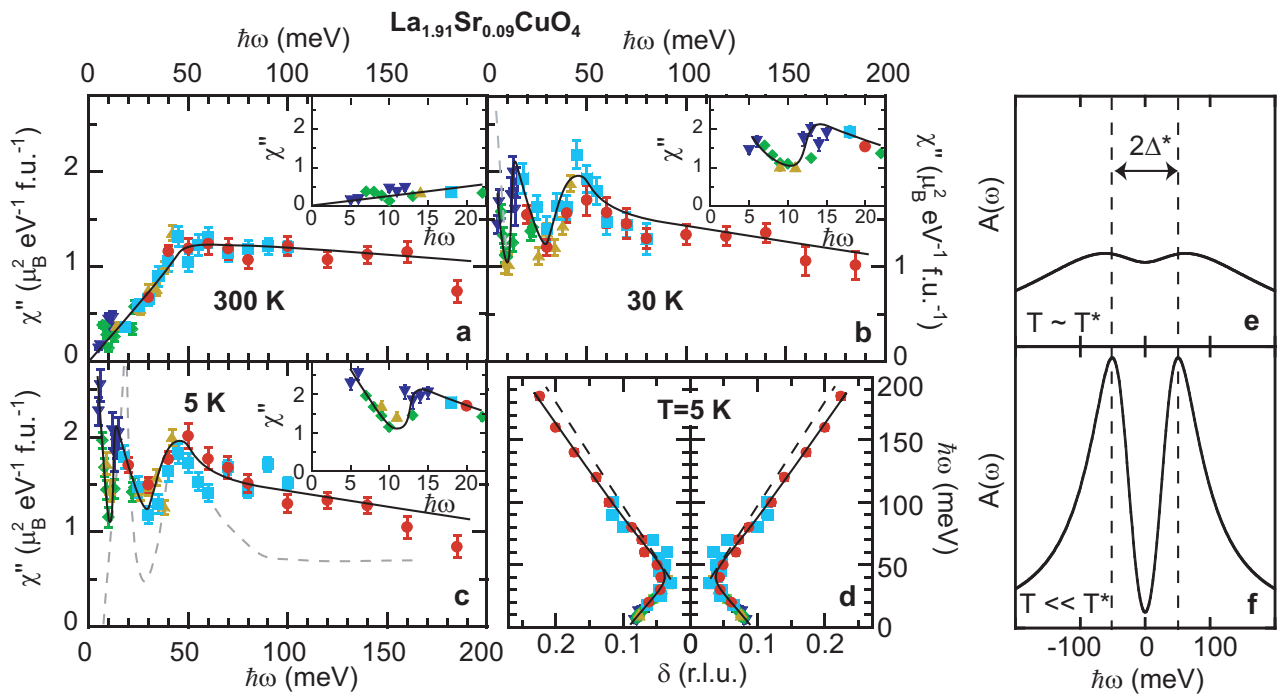}
\end{center}
\caption{(Color online) (a)-(c) Local or $\mathbf{q}$-integrated susceptibility $\chi^{\prime\prime}(\omega)$ extracted from fitting Eq.~\ref{Eq:Sato_Maki} to our complete dataset.  Insets are expanded views of the low-energy region. Symbols indicate $E_{i}$: 40 ($\blacktriangledown$), 55 ($\blacklozenge$), 90 ($\blacktriangle$), 160 ({\scriptsize $\blacksquare$}), 240 meV ({\large $\bullet$}). Solid lines are guides to the eye.  The grey dashed line in (c) is $\chi^{\prime\prime}(\omega)$ for LSCO ($x$=0.16) \cite{Vignolle2007a}. (d) Incommensurability parameter $\delta(\omega)$ for 5~K (30~K is indistinguishable), dashed line is the trend line for 300~K. (e)-(f) Plots of the symmetrized ARPES spectral function $A(\omega)$, (e) at high temperature and (f) deep in the pseudogap state (based on Ref. \cite{Norman1998a}).
}
\label{Fig:params}
\end{figure*}

Fig.~\ref{Fig:params} summarizes the main findings of this work. Panels \ref{Fig:params}(a)-(c) show the evolution of the local susceptibility with temperature. At $T=300$~K [Fig.~~\ref{Fig:params}(a)], we observe a heavily damped response with a characteristic energy scale of about 50~meV.  Interestingly, the response is roughly that of the marginal Fermi liquid \cite{Varma1989a} form, $\chi^{\prime\prime}(\omega) \sim \tanh(\omega/T)$, postulated to account for the strong temperature dependence of the electronic properties of the high-$T_c$ superconductors, and previously observed over a much smaller energy range in very underdoped La$_{2-x}$(Ba,Sr)$_{x}$CuO$_4$ \cite{Keimer1991,Hayden1991a}. The lack of structure in $\chi^{\prime\prime}(\omega)$ below 50~meV shows that the spin excitations are strongly damped at 300~K. On lowering the temperature to $T=30$~K [Fig.~~\ref{Fig:params}(b)] i.e. just above $T_c$=22~K, we observe a dramatic change in the response below about 70~meV. $\chi^{\prime\prime}(\omega)$ becomes more structured with peaks developing at 15 and 45~meV and an upturn at low energies below 10 meV. The upturn below 10 meV is most likely due to the freezing of the low-frequency fluctuations generally seen for lightly doped La$_{2-x}$Sr$_{x}$CuO$_{4}$ samples \cite{Matsushita1999a,Julien2003}. For $\hbar\omega > $\ 10~meV, we observe the double-peaked structure seen at optimal doping \cite{Vignolle2007a}, with peaks at approximately 15 and 50 meV (compared to 18 and 50 meV for $x$=0.16). The double-peaked structure in $\chi^{\prime\prime}(\omega)$ is less pronounced for underdoped LSCO ($x=0.085$) than for optimally doped LSCO ($x=0.16$) [see the solid and dashed lines in Fig.~3(c)]. There is a weaker dip near 25 meV and more spectral weight above 50 meV making the response more like the approximately flat $\chi^{\prime\prime}(\omega)$ observed in antiferromagnetic La$_{2}$CuO$_{4}$ \cite{Hayden1996a}. On lowering the temperature to $T=$\ 5~K [Fig.~~\ref{Fig:params}(c)], i.e. below $T_c$, we observe little change in the magnetic response except for a small increase in $\chi^{\prime\prime}(\omega)$ below 10 meV due to spin freezing which is consistent with other data on underdoped LSCO \cite{Lee2000a,Chang2007a,Hayden1991a}.

Our data also give information about the dispersion of the magnetic excitations with energy.  Fig.~\ref{Fig:slices}(a)-(d) show maps of $\chi^{\prime\prime}(\mathbf{q},\omega)$ for $T=$\ 5~K at various energies. We observe the same rotation of the four-peaked pattern with energy that has been seen in YBa$_{2}$Cu$_{3}$O$_{6.6}$ \cite {Hayden2004} and possibly seen in optimally doped La$_{2-x}$Sr$_{x}$CuO$_{4}$ \cite{Vignolle2007a}. The high-energy pattern [Fig.~\ref{Fig:slices}(d)] appears to be more anisotropic than in optimally doped LSCO \cite{Vignolle2007a}. Fig.~\ref{Fig:params}(d) shows the dispersion of $\delta(\omega)$ extracted from our fitting analysis. The form of $\delta(\omega)$ is very similar to that observed at optimal doping with a minimum in $|\delta|$ at about 50 meV.

A striking feature of the present data is the evolution of $\chi^{\prime\prime}(\omega)$ with temperature between 300~K and 30~K. As mentioned in the introduction, many charge spectroscopies show dramatic changes as the pseudogap state develops. The present results are unusual in that instead of a `spin pseudogap' forming over a wide energy range, additional spectral weight appears at lower energies (below about 70 meV) as we move into the pseudogap state.  It is known that ARPES shows a momentum dependent pseudogap well above $T_c$ in underdoped high-$T_c$ superconductors \cite{Norman1998}.  Recent ARPES measurements on underdoped La$_{2-x}$Sr$_{x}$CuO$_{4}$ ($x=0.105$) show a pseudogap of $2\Delta^{*}=$\ 51~meV for $T \gtrsim T_c$ \cite{Shi2008a}. We expect $2\Delta^{*}$ be slightly higher for our sample because it is more underdoped. Thus the pseudogap (determined from ARPES) appears to determine approximately the region where the structured spin excitations develop as the temperature is lowered from 300 to 30~K.

We now consider two general scenarios that could explain the emergence of the observed structured and coherent excitations as the pseudogap develops: (i) the concomitant appearance of some form of magnetic order; (ii) a reduction of the damping or scattering. The first scenario seems the least likely because only very weak spin freezing or magnetic order is observed for the compositions studied here \cite{Matsushita1999a}. Yet the observed $\chi^{\prime\prime}(\omega)$ has a magnitude comparable with the parent antiferromagnet La$_{2}$CuO$_{4}$ \cite{Hayden1996a} which has an ordered moment $\approx$ 0.6~$\mu_B$.  The second scenario involves a reduction of the electronic damping on entering the pseudogap state. This picture fits particulary well with ARPES measurements which have been interpreted using self energies with strongly temperature-dependent damping \cite{Norman1998a,Chubukov2007a}. In such pictures the pseudogap $\Delta^{\star}$ is a remnant of the superconducting gap.  At high temperatures it is heavily damped. As the temperature is lowered the damping is reduced and the effects of the pseudogap become visible over an energy scale $2\Delta^{\star}$ (see Fig.~\ref{Fig:params} for a schematic picture). This picture may provide a basis for explaining our results since ARPES and other measurements suggest $2\Delta^{*} \gtrsim $\ 50~meV for the La$_{2-x}$Sr$_{x}$CuO$_{4}$ ($x=0.085$) studied here and we observed large changes over a similar energy range $E \lesssim$\ 70~meV.

In conclusion, we have shown that highly structured and coherent magnetic excitations exist over a wide energy scale ($5<E<200$\ meV) in underdoped and superconducting La$_{2-x}$Sr$_{x}$CuO$_{4}$. At high energies, $E \gtrsim 60$\ 60~meV, the excitations have a similar dispersion to pure La$_{2}$CuO$_{4}$, albeit with a slightly reduced intensity. At low energies, $E<60$\ meV, we find a double-peaked structure in $\chi^{\prime\prime}(\omega)$ and a four-peak structure in wavevector which develop as the temperature is lowered from 300 K to 30 K i.e. as we move into the pseudogap region of the phase diagram. We note that the energy scale over which $\chi^{\prime\prime}(\mathbf{q},\omega)$ changes most corresponds approximately to the pseudogap energy $2 \Delta^{\star}$ for our sample. This is consistent with a strong reduction in damping of the magnetic excitations as the pseudogap develops.

We thank A. V. Chubukov and M. R. Norman for useful discussions.


\begin{thebibliography}{26}
\providecommand{\natexlab}[1]{#1}
\providecommand{\bibnamefont}[1]{#1}
\providecommand{\bibfnamefont}[1]{#1}
\providecommand{\citenamefont}[1]{#1}
\providecommand{\url}[1]{\texttt{#1}}
\providecommand{\urlprefix}{URL }
\providecommand{\bibinfo}[2]{#2}
\providecommand{\eprint}[2][]{\url{#2}}

\bibitem[{\citenamefont{Timusk and Statt}(1999)}]{Timusk1999}
\bibinfo{author}{\bibfnamefont{T.}~\bibnamefont{Timusk}} \bibnamefont{and}
  \bibinfo{author}{\bibfnamefont{B.}~\bibnamefont{Statt}},
  \bibinfo{journal}{Rep. Prog. Phys.} \textbf{\bibinfo{volume}{62}},
  \bibinfo{pages}{61} (\bibinfo{year}{1999}).

\bibitem[{\citenamefont{Tallon and Loram}(2001)}]{Tallon2001}
\bibinfo{author}{\bibfnamefont{J.~L.} \bibnamefont{Tallon}} \bibnamefont{and}
  \bibinfo{author}{\bibfnamefont{J.~W.} \bibnamefont{Loram}},
  \bibinfo{journal}{Physica C} \textbf{\bibinfo{volume}{349}},
  \bibinfo{pages}{53} (\bibinfo{year}{2001}).

\bibitem[{\citenamefont{Kyung} \emph{et~al.}(2006)}]{Kyung2006a}
\bibinfo{author}{\bibfnamefont{B.}~\bibnamefont{Kyung}}, \emph{et~al.},
  \bibinfo{journal}{Phys. Rev. B} \textbf{\bibinfo{volume}{73}},
  \bibinfo{pages}{165114} (\bibinfo{year}{2006}).

\bibitem[{\citenamefont{Varma and Zhu}(2007)}]{Varma2007a}
\bibinfo{author}{\bibfnamefont{C.~M.} \bibnamefont{Varma}} \bibnamefont{and}
  \bibinfo{author}{\bibfnamefont{L.}~\bibnamefont{Zhu}},
  \bibinfo{journal}{Phys. Rev. Lett.} \textbf{\bibinfo{volume}{98}},
  \bibinfo{pages}{177004} (\bibinfo{year}{2007}).

\bibitem[{\citenamefont{Schmalian} \emph{et~al.}(1998)\citenamefont{Schmalian,
  Pines, and Stojkovic}}]{Schmalian2008a}
\bibinfo{author}{\bibfnamefont{J.}~\bibnamefont{Schmalian}},
  \bibinfo{author}{\bibfnamefont{D.}~\bibnamefont{Pines}}, \bibnamefont{and}
  \bibinfo{author}{\bibfnamefont{B.}~\bibnamefont{Stojkovic}},
  \bibinfo{journal}{Phys. Rev. Lett.} \textbf{\bibinfo{volume}{80}},
  \bibinfo{pages}{3839} (\bibinfo{year}{1998}).

\bibitem[{\citenamefont{Emery and Kivelson}(1995)}]{Emery1995a}
\bibinfo{author}{\bibfnamefont{V.~J.} \bibnamefont{Emery}} \bibnamefont{and}
  \bibinfo{author}{\bibfnamefont{S.~A.} \bibnamefont{Kivelson}},
  \bibinfo{journal}{Nature} \textbf{\bibinfo{volume}{374}},
  \bibinfo{pages}{434} (\bibinfo{year}{1995}).

\bibitem[{\citenamefont{Norman}
  \emph{et~al.}(1998{\natexlab{a}})\citenamefont{Norman, Randeria, Ding, and
  Campuzano}}]{Norman1998a}
\bibinfo{author}{\bibfnamefont{M.~R.} \bibnamefont{Norman}},
  \bibinfo{author}{\bibfnamefont{M.}~\bibnamefont{Randeria}},
  \bibinfo{author}{\bibfnamefont{H.}~\bibnamefont{Ding}}, \bibnamefont{and}
  \bibinfo{author}{\bibfnamefont{J.~C.} \bibnamefont{Campuzano}},
  \bibinfo{journal}{Phys. Rev. B} \textbf{\bibinfo{volume}{57}},
  \bibinfo{pages}{11093} (\bibinfo{year}{1998}{\natexlab{a}}).

\bibitem[{\citenamefont{Chubukov} \emph{et~al.}(2007)\citenamefont{Chubukov,
  Norman, Millis, and Abrahams}}]{Chubukov2007a}
\bibinfo{author}{\bibfnamefont{A.~V.} \bibnamefont{Chubukov}},
  \bibinfo{author}{\bibfnamefont{M.~R.} \bibnamefont{Norman}},
  \bibinfo{author}{\bibfnamefont{A.~J.} \bibnamefont{Millis}},
  \bibnamefont{and} \bibinfo{author}{\bibfnamefont{E.}~\bibnamefont{Abrahams}},
  \bibinfo{journal}{Phys. Rev. B} \textbf{\bibinfo{volume}{76}},
  \bibinfo{pages}{180501} (\bibinfo{year}{2007}).

\bibitem[{\citenamefont{Bourges} \emph{et~al.}(1997)}]{Bourges1997}
\bibinfo{author}{\bibfnamefont{P.}~\bibnamefont{Bourges}}, \emph{et~al.},
  \bibinfo{journal}{Phys. Rev. B} \textbf{\bibinfo{volume}{56}},
  \bibinfo{pages}{11439} (\bibinfo{year}{1997}).

\bibitem[{\citenamefont{Dai} \emph{et~al.}(1999)}]{Dai1999}
\bibinfo{author}{\bibfnamefont{P.~C.} \bibnamefont{Dai}}, \emph{et~al.},
  \bibinfo{journal}{Science} \textbf{\bibinfo{volume}{284}},
  \bibinfo{pages}{1344} (\bibinfo{year}{1999}).

\bibitem[{\citenamefont{Keimer} \emph{et~al.}(1991)}]{Keimer1991}
\bibinfo{author}{\bibfnamefont{B.}~\bibnamefont{Keimer}}, \emph{et~al.},
  \bibinfo{journal}{Phys. Rev. Lett.} \textbf{\bibinfo{volume}{67}},
  \bibinfo{pages}{1930} (\bibinfo{year}{1991}).

\bibitem[{\citenamefont{Hayden} \emph{et~al.}(1991)}]{Hayden1991a}
\bibinfo{author}{\bibfnamefont{S.~M.} \bibnamefont{Hayden}}, \emph{et~al.},
  \bibinfo{journal}{Phys. Rev. Lett.} \textbf{\bibinfo{volume}{66}},
  \bibinfo{pages}{821} (\bibinfo{year}{1991}).

\bibitem[{\citenamefont{Lee} \emph{et~al.}(2000)}]{Lee2000a}
\bibinfo{author}{\bibfnamefont{C.-H.} \bibnamefont{Lee}}, \emph{et~al.},
  \bibinfo{journal}{J. Phys. Soc. Jap.} \textbf{\bibinfo{volume}{69}},
  \bibinfo{pages}{1170} (\bibinfo{year}{2000}).

\bibitem[{\citenamefont{Kofu} \emph{et~al.}(2007)\citenamefont{Kofu, Yokoo,
  Trouw, and Yamada}}]{Kofu2007a}
\bibinfo{author}{\bibfnamefont{M.}~\bibnamefont{Kofu}},
  \bibinfo{author}{\bibfnamefont{T.}~\bibnamefont{Yokoo}},
  \bibinfo{author}{\bibfnamefont{F.}~\bibnamefont{Trouw}}, \bibnamefont{and}
  \bibinfo{author}{\bibfnamefont{K.}~\bibnamefont{Yamada}}
  (\bibinfo{year}{2007}), \bibinfo{note}{arXiv:0710.5766}.

\bibitem[{\citenamefont{Komiya} \emph{et~al.}(2002)\citenamefont{Komiya, Ando,
  Sun, and Lavrov}}]{Komiya2002a}
\bibinfo{author}{\bibfnamefont{S.}~\bibnamefont{Komiya}},
  \bibinfo{author}{\bibfnamefont{Y.}~\bibnamefont{Ando}},
  \bibinfo{author}{\bibfnamefont{X.~F.} \bibnamefont{Sun}}, \bibnamefont{and}
  \bibinfo{author}{\bibfnamefont{A.~N.} \bibnamefont{Lavrov}},
  \bibinfo{journal}{Phys. Rev. B} \textbf{\bibinfo{volume}{65}},
  \bibinfo{pages}{214535} (\bibinfo{year}{2002}).

\bibitem[{\citenamefont{Lipscombe} \emph{et~al.}(2007)\citenamefont{Lipscombe,
  Hayden, Vignolle, McMorrow, and Perring}}]{Lipscombe2007a}
\bibinfo{author}{\bibfnamefont{O.~J.} \bibnamefont{Lipscombe}},
  \bibinfo{author}{\bibfnamefont{S.~M.} \bibnamefont{Hayden}},
  \bibinfo{author}{\bibfnamefont{B.}~\bibnamefont{Vignolle}},
  \bibinfo{author}{\bibfnamefont{D.~F.} \bibnamefont{McMorrow}},
  \bibnamefont{and} \bibinfo{author}{\bibfnamefont{T.~G.}
  \bibnamefont{Perring}}, \bibinfo{journal}{Phys. Rev. Lett.}
  \textbf{\bibinfo{volume}{99}}, \bibinfo{pages}{067002}
  (\bibinfo{year}{2007}).

\bibitem[{\citenamefont{Cheong} \emph{et~al.}(1991)}]{Cheong1991}
\bibinfo{author}{\bibfnamefont{S.~W.} \bibnamefont{Cheong}}, \emph{et~al.},
  \bibinfo{journal}{Phys. Rev. Lett.} \textbf{\bibinfo{volume}{67}},
  \bibinfo{pages}{1791} (\bibinfo{year}{1991}).

\bibitem[{\citenamefont{Vignolle} \emph{et~al.}(2007)}]{Vignolle2007a}
\bibinfo{author}{\bibfnamefont{B.}~\bibnamefont{Vignolle}}, \emph{et~al.},
  \bibinfo{journal}{Nat. Phys.} \textbf{\bibinfo{volume}{3}},
  \bibinfo{pages}{163} (\bibinfo{year}{2007}).

\bibitem[{\citenamefont{Varma} \emph{et~al.}(1989)\citenamefont{Varma,
  Littlewood, Schmitt-Rink, Abrahams, and Ruckenstein}}]{Varma1989a}
\bibinfo{author}{\bibfnamefont{C.~M.} \bibnamefont{Varma}},
  \bibinfo{author}{\bibfnamefont{P.~B.} \bibnamefont{Littlewood}},
  \bibinfo{author}{\bibfnamefont{S.}~\bibnamefont{Schmitt-Rink}},
  \bibinfo{author}{\bibfnamefont{E.}~\bibnamefont{Abrahams}}, \bibnamefont{and}
  \bibinfo{author}{\bibfnamefont{A.~E.} \bibnamefont{Ruckenstein}},
  \bibinfo{journal}{Phys. Rev. Lett.} \textbf{\bibinfo{volume}{63}},
  \bibinfo{pages}{1996} (\bibinfo{year}{1989}).

\bibitem[{\citenamefont{Matsushita} \emph{et~al.}(1999)}]{Matsushita1999a}
\bibinfo{author}{\bibfnamefont{H.}~\bibnamefont{Matsushita}}, \emph{et~al.},
  \bibinfo{journal}{J. Phys. Chem. Sol.} \textbf{\bibinfo{volume}{60}},
  \bibinfo{pages}{1071} (\bibinfo{year}{1999}).

\bibitem[{\citenamefont{Julien}(2003)}]{Julien2003}
\bibinfo{author}{\bibfnamefont{M.-H.} \bibnamefont{Julien}},
  \bibinfo{journal}{Physica B} \textbf{\bibinfo{volume}{329–-333}},
  \bibinfo{pages}{693} (\bibinfo{year}{2003}).

\bibitem[{\citenamefont{Hayden} \emph{et~al.}(1996)}]{Hayden1996a}
\bibinfo{author}{\bibfnamefont{S.~M.} \bibnamefont{Hayden}}, \emph{et~al.},
  \bibinfo{journal}{Phys. Rev. Lett.} \textbf{\bibinfo{volume}{76}},
  \bibinfo{pages}{1344} (\bibinfo{year}{1996}).

\bibitem[{\citenamefont{Chang} \emph{et~al.}(2007)}]{Chang2007a}
\bibinfo{author}{\bibfnamefont{J.}~\bibnamefont{Chang}}, \emph{et~al.},
  \bibinfo{journal}{Phys. Rev. Lett.} \textbf{\bibinfo{volume}{98}},
  \bibinfo{pages}{077004} (\bibinfo{year}{2007}).

\bibitem[{\citenamefont{Hayden} \emph{et~al.}(2004)\citenamefont{Hayden, Mook,
  Dai, Perring, and Dogan}}]{Hayden2004}
\bibinfo{author}{\bibfnamefont{S.~M.} \bibnamefont{Hayden}},
  \bibinfo{author}{\bibfnamefont{H.~A.} \bibnamefont{Mook}},
  \bibinfo{author}{\bibfnamefont{P.~C.} \bibnamefont{Dai}},
  \bibinfo{author}{\bibfnamefont{T.~G.} \bibnamefont{Perring}},
  \bibnamefont{and} \bibinfo{author}{\bibfnamefont{F.}~\bibnamefont{Dogan}},
  \bibinfo{journal}{Nature} \textbf{\bibinfo{volume}{429}},
  \bibinfo{pages}{531} (\bibinfo{year}{2004}).

\bibitem[{\citenamefont{Norman} \emph{et~al.}(1998{\natexlab{b}})}]{Norman1998}
\bibinfo{author}{\bibfnamefont{M.~R.} \bibnamefont{Norman}}, \emph{et~al.},
  \bibinfo{journal}{Nature} \textbf{\bibinfo{volume}{392}},
  \bibinfo{pages}{157} (\bibinfo{year}{1998}{\natexlab{b}}).

\bibitem[{\citenamefont{Shi} \emph{et~al.}(2008)}]{Shi2008a}
\bibinfo{author}{\bibfnamefont{M.}~\bibnamefont{Shi}}, \emph{et~al.},
  \bibinfo{journal}{Phys. Rev. Lett.} \textbf{\bibinfo{volume}{101}},
  \bibinfo{pages}{047002} (\bibinfo{year}{2008}).

\end{thebibliography}

\end{document}